\documentclass{appolb}
\usepackage{epsfig}
\usepackage{cite}

\begin{document}
\title{DIRECT DETERMINATION OF THE TOTAL WIDTH OF THE $\eta'$ MESON USING\\THE COSY--11 APPARATUS%
\thanks{Presented at the Symposium on Meson Physics, Cracow, 01-04 October 2008.}%
}
\author{Eryk Czerwi{\'n}ski on behalf the COSY--11 collaboration
        \address{\begin{center}Institute of Physics, Jagiellonian University, 30-059 Cracow, Poland\\
                 \&\\
                 Institute for Nuclear Physics and J{\"u}lich Center for Hadron Physics,\\
                 Research Center J{\"u}lich, D-52425 J{\"u}lich, Germany\end{center}
                }
       }
\maketitle
\begin{abstract}
We describe the experimental method and consecutive
steps of the analysis of the $pp\to pp\eta'$ reaction 
measured by means of the COSY--11 detection setup. 
The conducted investigation aim at the 
determination of the total width of the $\eta'$ 
meson directly from its mass distribution.
The preliminary results show that the statistical error is in the order of $\approx10$~keV.
\end{abstract}
\PACS{13.75.Cs, 14.40.Aq}

\section{Introduction}
Inputs for the phenomenological desription of Quantum Chromo-Dy\-na\-mics in the non-perturbative regime~\cite{bugra}
can be provided from studies of the $\eta'$ meson production~\cite{hab} and its decays~\cite{andrzej}.
Such investigations are of interest on its own account and are planned
to be conducted with
WASA-at-COSY~\cite{hhadam}, KLOE-2~\cite{ambrosino}
and \mbox{CBall-at-MAMI~\cite{thomas}} experiments.
In addition, precise determinations of the partial widths for the $\eta^{\prime}$ decay
channels will be helpful for the development of the Chiral Perturbation Theory
as 
constrains
for the calculations.
However, regardless of the branching ratios of the $\eta'$ meson decays which are typically known with
an accuracy better than 1.5\%, the total width is established
about 10 times less accurate~\cite{pdg} and cause that 
the experimental precision of the partial width for various decay channels
 -- where only the branching ratio is known or will be measured --
is governed by  the precision in the knowledge of the total width.

In the last issue of the Review of Particle Physics
a value of (0.205\-$\pm$0.015)~MeV/c$^{2}$ is given for
the total width of the $\eta'$ meson ($\Gamma_{\eta'}$) resulting from a fit
which includes a 
combination of partial widths obtained from
integrated cross sections and branching ratios from 50 measurements~\cite{pdg}.

This indirect determination introduces correlation between the value of the $\Gamma_{\eta'}$ and the branching ratios
making difficulties in the investigations of the other properties of the $\eta'$ meson~\cite{biagio}.

On the other hand, the mean value from the only two
direct measurements~\cite{binnie,wurzinger} taken into account by PDG
amounts to \mbox{(0.30$\pm$0.09)~MeV/c$^{2}$}~\cite{pdg} and differs 
from the fit result for $\Gamma_{\eta'}$.

Therefore, a precise determination of the natural width of the $\eta'$ meson  may have
a large impact on the physics results which will be derived from
measurements at facilities like
COSY, DA$\Phi$NE-2 and MAMI-C and may solve the discrepancy in the values
from the direct measurements and the indirect determination which is 
recommended as the nominal width.

This situation encouraged us to conduct investigations of
$\Gamma_{\eta'}$ directly from
the missing mass distribution of the $pp \to pp\eta^{\prime}$
reaction measured near
the kinematical threshold.  The advantage of a study close to the threshold
is that the uncertainties of the missing mass determination
are considerably reduced since at threshold the value of $\partial(mm)\slash\partial p$ approach zero
(\emph{mm} = missing mass, \emph{p} = momentum of the outgoing protons).

\section{Experiment}
\begin{figure}[!t]
 \parbox{0.60\textwidth}{
    \vspace{-1cm}
    \epsfig{file=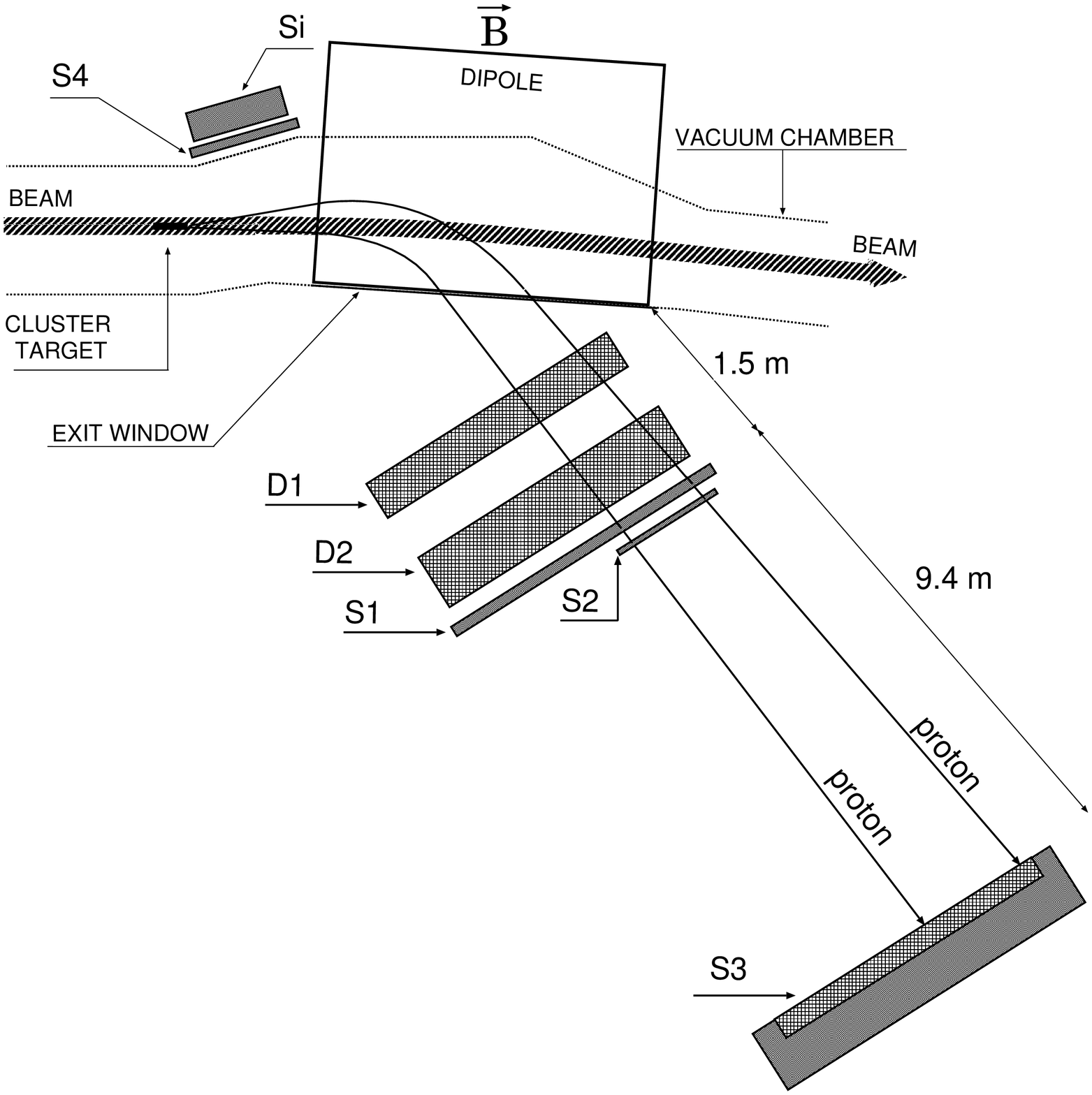,width=0.59\textwidth,angle=0}
}
 \parbox{0.40\textwidth}{
  \epsfig{file=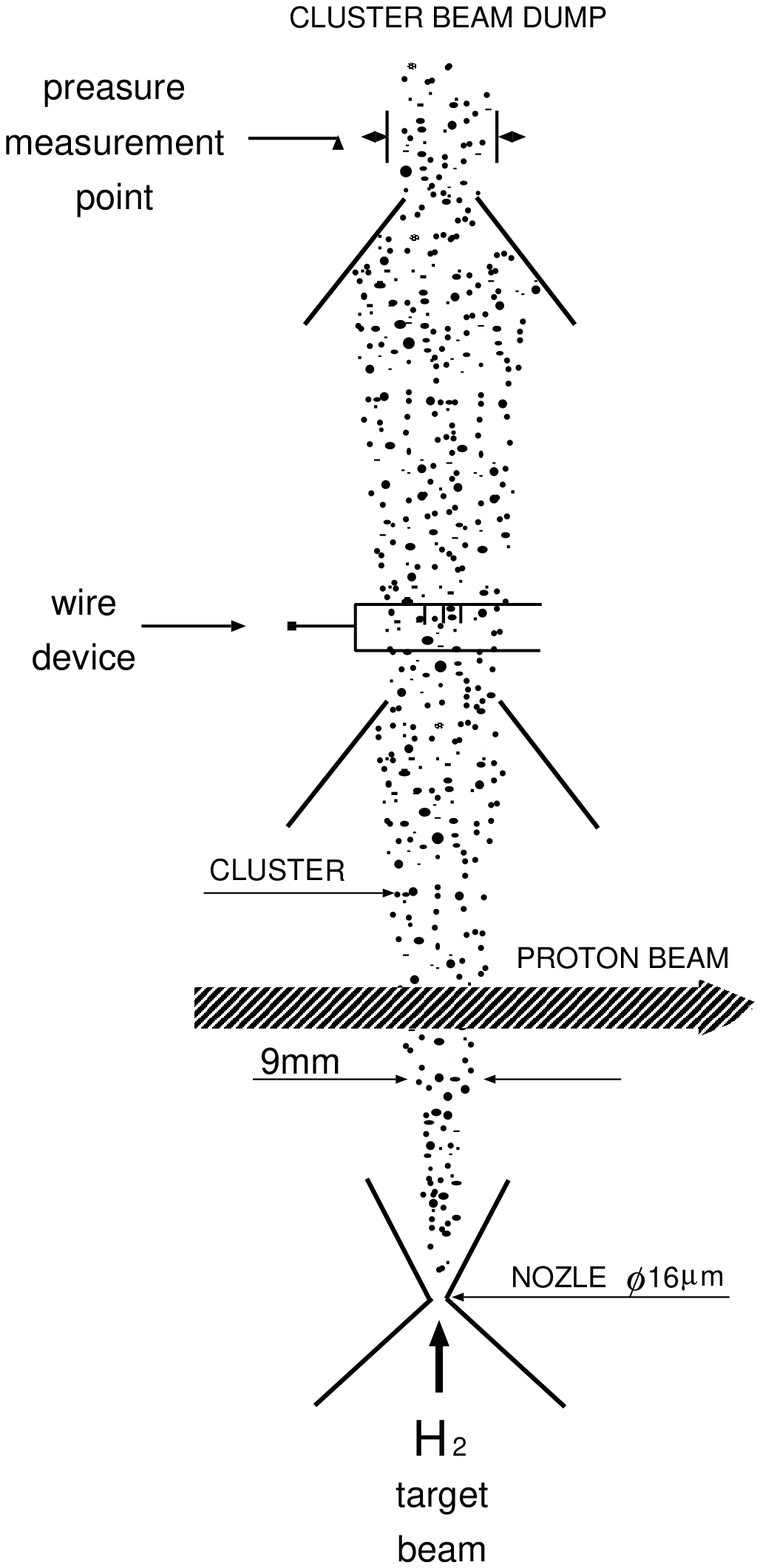,height=0.45\textheight,width=0.39\textwidth,angle=0}
  }
 \parbox{0.40\textwidth}{
    \vspace{-4cm}
    \epsfig{file=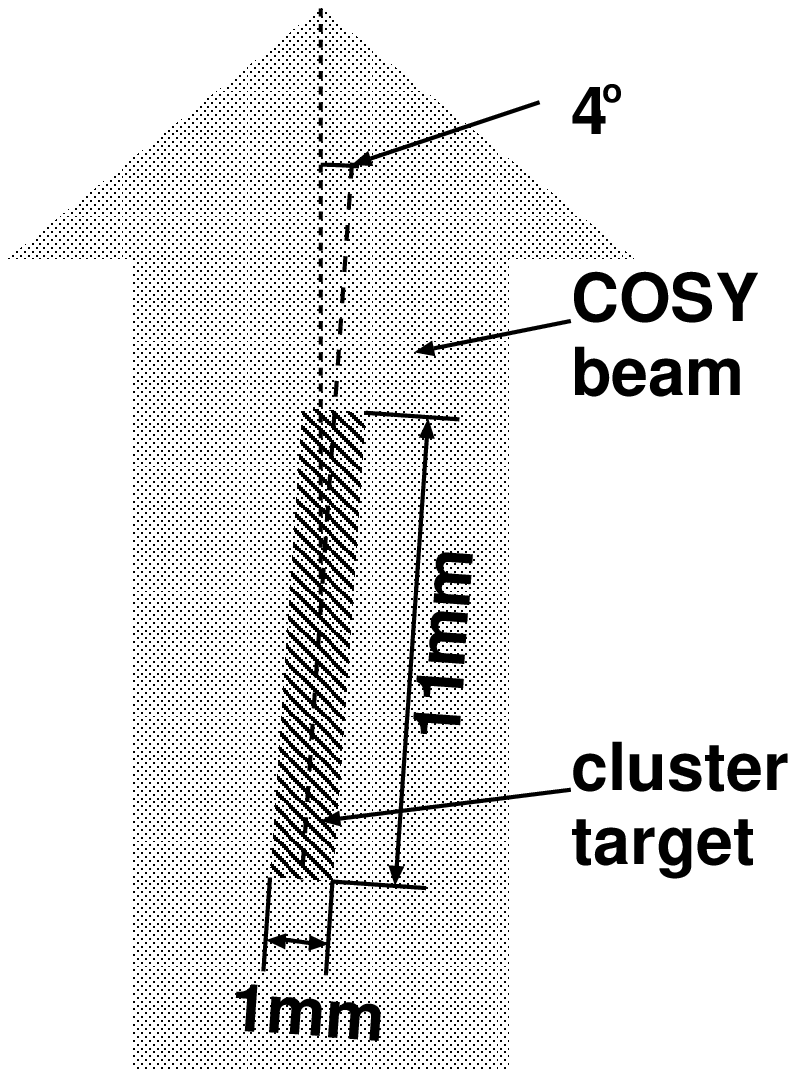,width=0.20\textwidth,angle=0}
 }
 \caption{\small{{\bf Left:}~Top view at the COSY--11 detection setup with drift chambers (D1, D2) used
           for the reconstruction of trajectories of positively charged ejectiles
           and scintillator hodoscopes (S1, S2, S3) for the time of flight determination.
           The silicon pad (Si) and scintillator (S4) detectors
           register the elastically scattered protons used for
           monitoring purposes.
         {\bf Right:}~Schematic side view of the target and beam crossing. The diagnosis
           unit for the target dimensions measurement installed above the reaction region
           does not influence the $pp\to pp\eta'$ measurement.
         {\bf Bottom Left:}~Cross section of the cluster target stream in the COSY beam plane.
         }}
 \label{c11}
\end{figure}

In September and October 2006 data of the $pp\to ppX$ reaction were collected using the COSY--11
detection setup~\cite{brauksiepe}, a hydrogen cluster target~\cite{dombrowski} and the stochastically
cooled proton beam of COSY~\cite{prasuhn}. Five beam momenta: 3211, 3213, 3214, 3218 and 3224~MeV/c
(corresponding to the excess energy of 0.9, 1.5, 1.8, 3.1 and 5.0~MeV, respectively) were used.
The collision of a proton from the COSY beam with a cluster target proton may cause an $\eta'$ meson creation.
The ejected protons of the $pp\to pp\eta'$ reaction were separated from the circulating beam
by the magnetic field due to their lower momenta and were registered by the detection system consisting of drift chambers and scintillation
counters as depicted in Fig.~\ref{c11}~(top).
The reconstruction of the momentum vector for each registered particle is based on 
the measurement of the track direction by means of the drift chambers,
the knowledge of the dipole magnetic field and the target position.
Together with the independent determination of the particle velocity from the 
measured time of flight between the S1
and S3 scintillators the particle identification is provided.
The knowledge of the momenta of both protons
before and after the reaction allows to calculate the mass of a not observed particle or system of particles in the outgoing
channel, which in case of the $pp \to pp\eta'$ reaction should be equal to the mass of the $\eta'$ meson.

In order to improve the experimental resolution
of the four-momentum determination
and in order to decrease the momentum spread
of the beam protons reacting with the target
two major changes have been applied to the COSY-11 setup (Fig.~\ref{c11})
with respect to the previous measurements of the $pp\to pp\eta^{\prime}$ reactions~\cite{c11prl,c11pl,c11euro}.
Namely, the spatial resolution of the
drift chambers was improved by
increasing  the supply voltage up to  the maximum allowed value and also
the width of the target in the direction perpendicular to the COSY beam was decreased
from 9 to circa 1~mm~\cite{dombrowski,taeschner}.
%
\subsection{Detectors}
\begin{figure}[!b]
\vspace{-5mm}
 \begin{center}
  \epsfig{file=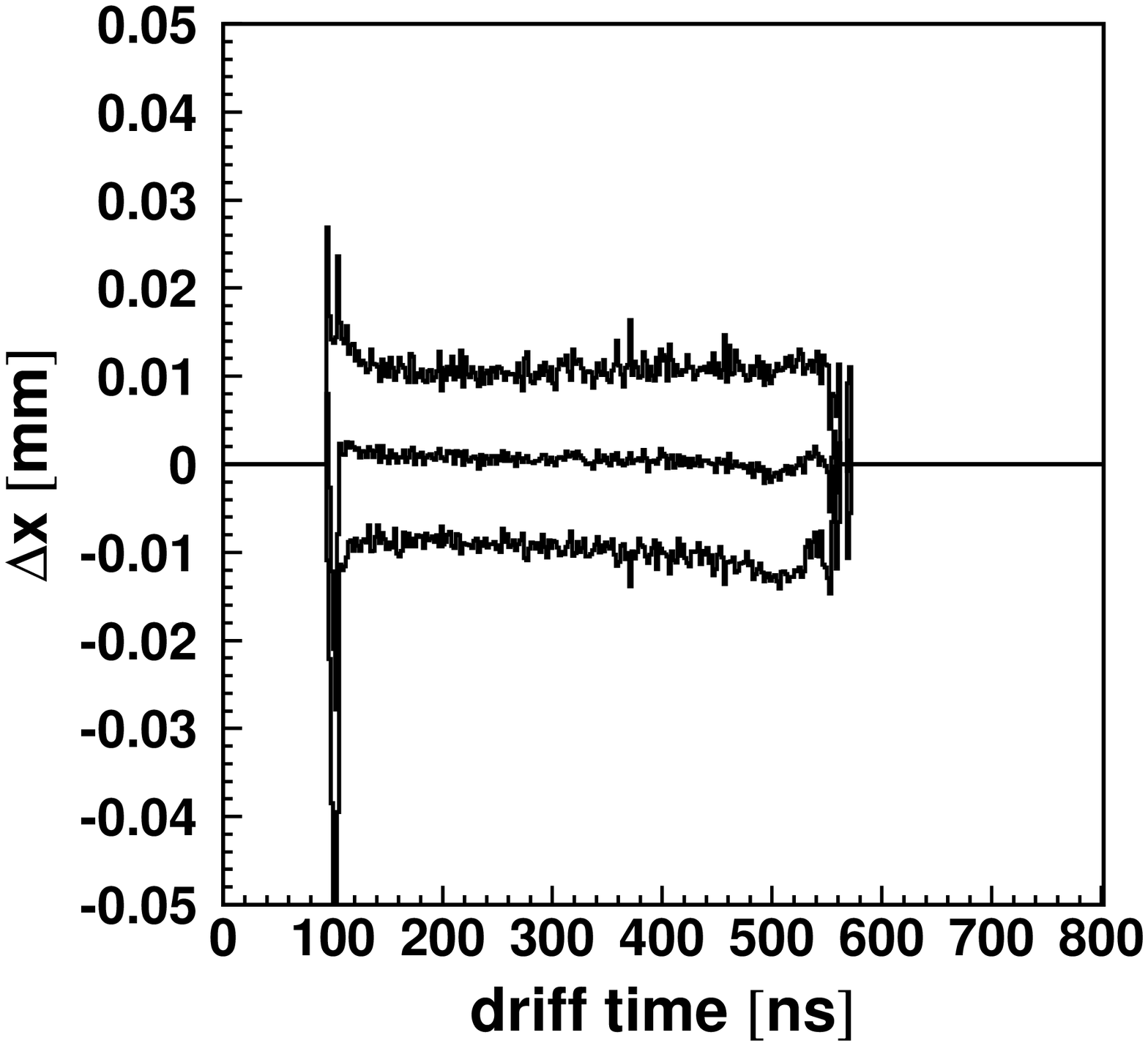,width=0.39\textwidth,angle=0}
  \epsfig{file=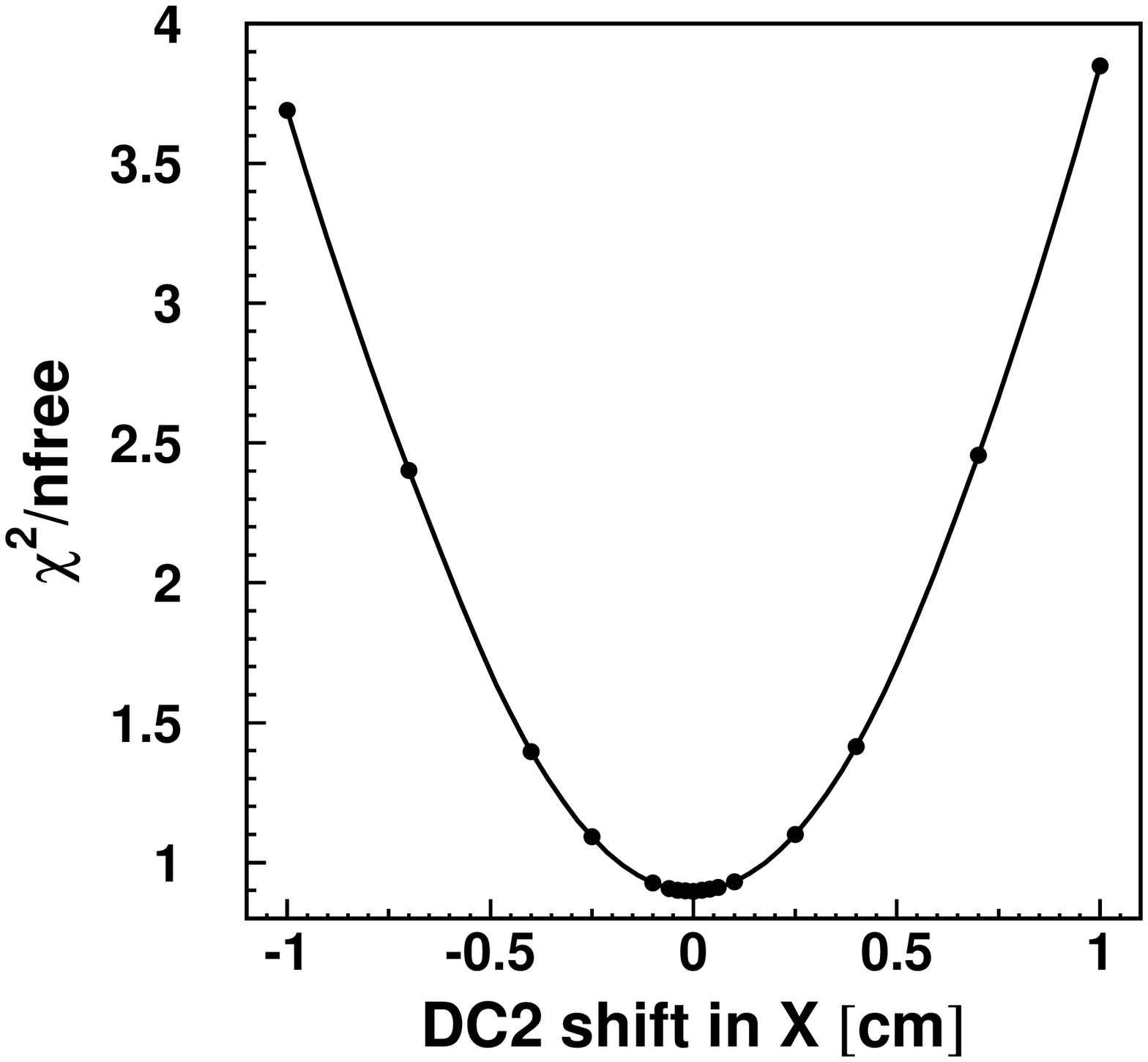,width=0.39\textwidth,angle=0}
  \epsfig{file=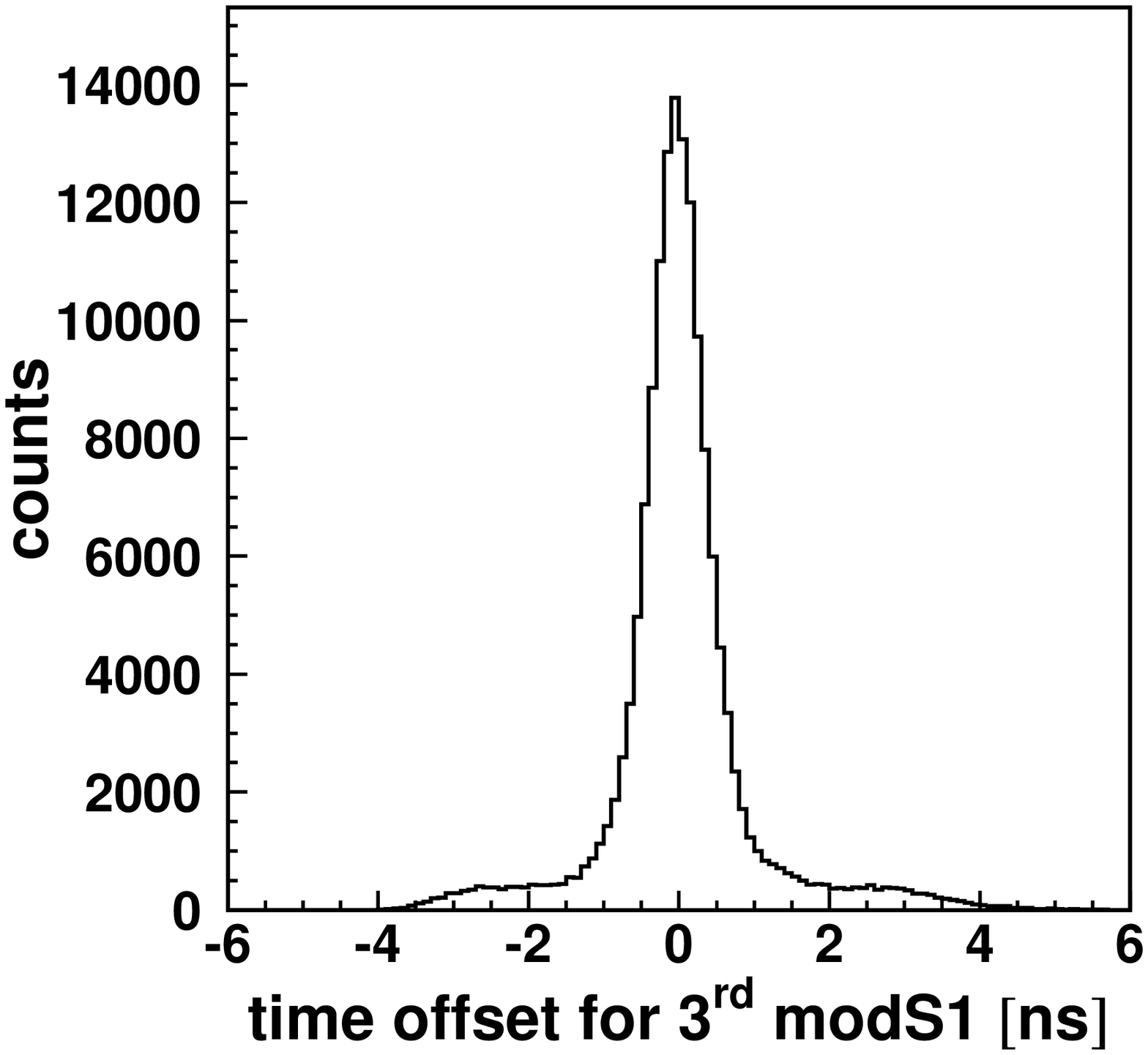,width=0.39\textwidth,angle=0}
  \epsfig{file=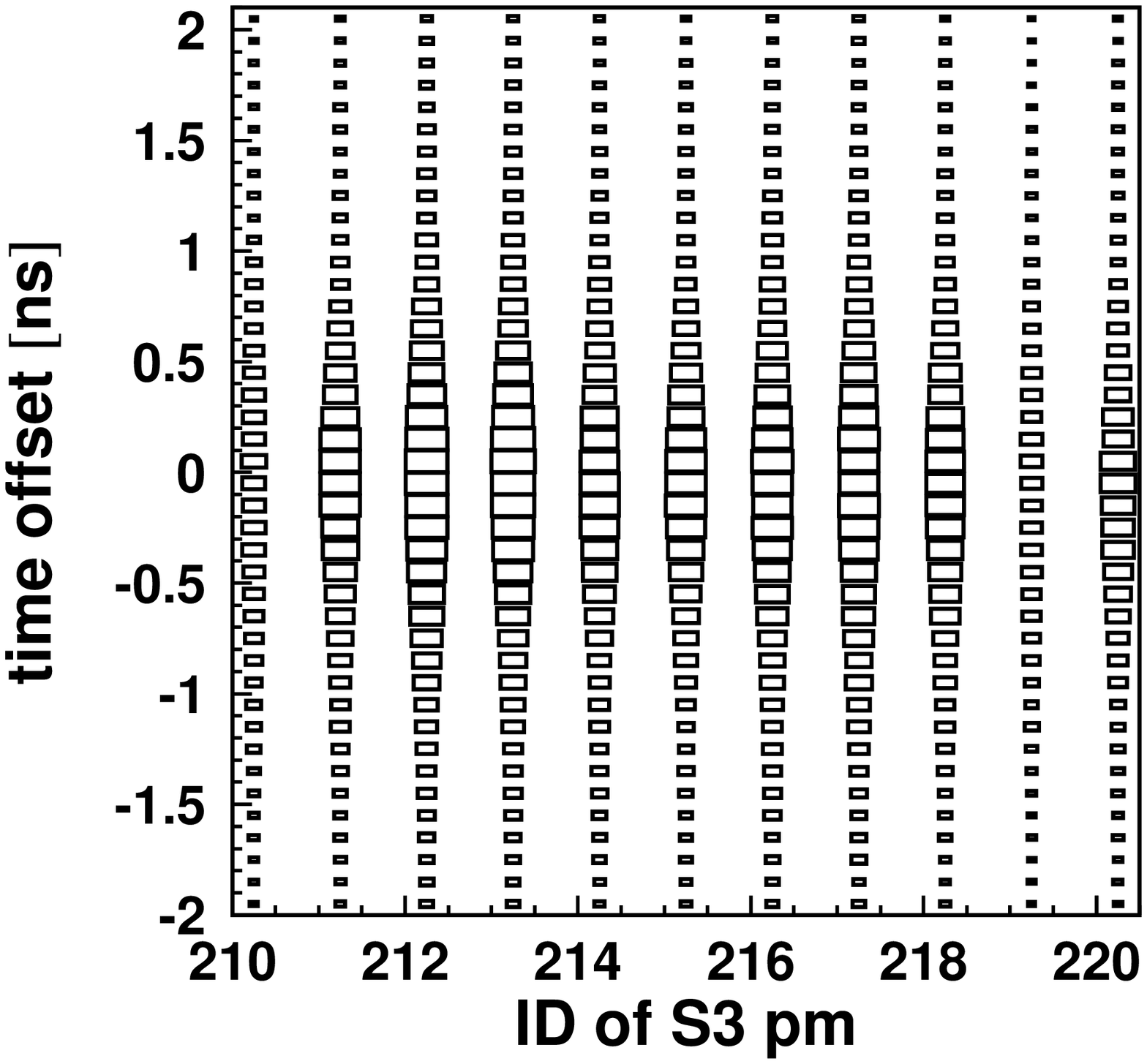,width=0.39\textwidth,angle=0}
 \end{center}
 \caption{\small{{\bf Top Left:}~~Average deviation ($\Delta X$) between 
           the measured and the fitted distances
           of tracks from the sense wire as a function of the drift time.
           The histogram around 0 corresponds to the average value
           of the $\Delta X$ distribution and the upper and lower lines denotes the 
           standard deviation of the $\Delta X$ distribution.
          {\bf Top Right:}~$\chi^2$ distribution for the fit of the straight line
           to the signals from both drift chambers as a function of the position
           of the second with respect to the first drift chamber.
          {\bf Bottom:}~Distributions of difference between 
           the time-of-flight measured using S1-S3 detectors
           and the time-of-flight calculated from the momentum 
           reconstructed based on the curvature of the trajectory in the magnetic field.
           As an example
           spectra for 3rd S1 module and a range of photomultipliers (pm) of S3 detector are shown.
           The counting rate of PM 210 and 219 is smaller since these photomultipliers
           are positioned at the edges of the detector.
         }}
 \label{detectors}
\end{figure}

The trajectory of a charged particle passing through the drift chambers (DC) is
obtained from the dependency between drift time of the electrons (from the gas ionized by that particle)
going towards the sense wire and the distance between sense wire and particle trajectory.
The evaluation of this dependency is based on the assumption that the particle trajectory inside
the drift chamber is a straight line.
The drift time to distance relation was calibrated for each 20 - 24 hours of the data
taking period in order to minimise fluctuations of the drift velocity
caused by variations of atmospheric pressure, air humidity
and gas mixture changes.
Figure~\ref{detectors} (upper left) illustrates that the obtained spatial resolution equals to about 100~$\mu$m.
 
The velocity of the charged particle is measured by means of the S1(S2) and S3 scintillator detectors applying
the time of flight (ToF) method. For the calibration of the scintillator detectors
we compare the time-of-flight
obtained from signals registered in the S1 and S3
detectors and the time-of-flight calculated from the reconstructed momentum of the particle.
As an example the lower plots in Fig.~\ref{detectors} 
present results of the calibration for  arbitrarily chosen photomultipliers (pm)
of the S1 and S3 detectors.

\begin{figure}[!b]
 \vspace{-7mm}
 \begin{center}
  \epsfig{file=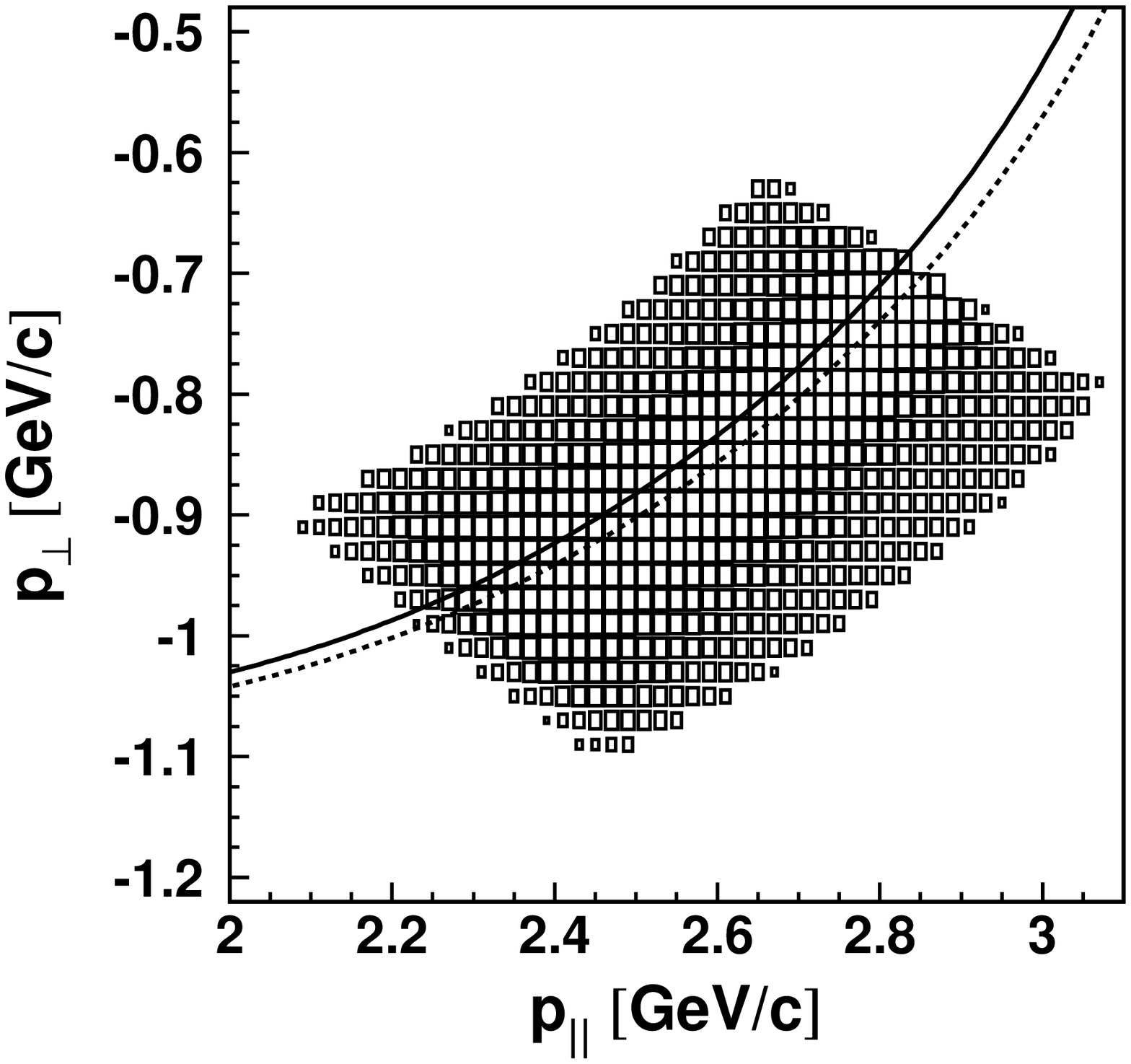,width=0.49\textwidth,angle=0}
  \epsfig{file=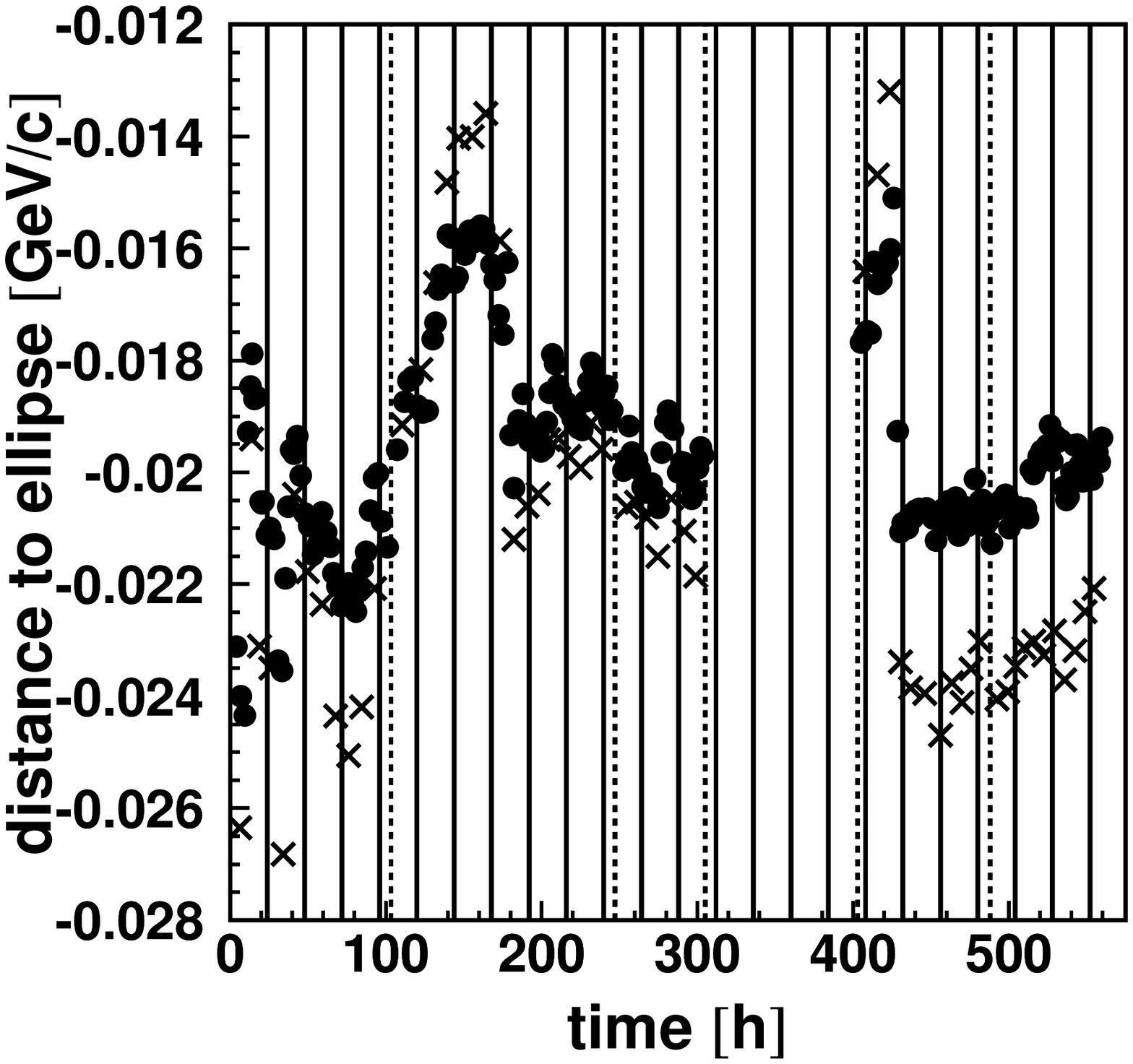,width=0.49\textwidth,angle=0}
 \end{center}
 \caption{\small{{\bf Left:}~Part of the kinematical ellipse of elastically scattered protons
           collected at a nominal beam momentum of 3211~MeV/c.
           The solid line denotes the position of the ellipse for the real beam momentum of 3211~MeV/c.
           The dashed line shows the centre of the reconstructed ellipse which corresponds to a beam momentum of 3250~MeV/c.
          {\bf Right:}~Distance of the reconstructed kinematical ellipse of the elastically scattered protons
           to the nominal (expected) one as a function of time of the measurement.
           The straight lines mark 24~hour intervals. The dashed lines separate
           the different beam momenta (3218, 3211, 3214, 3213 and 3224~MeV/c).
           Filled circles include events with a signal in the first S1 module,
           crosses for events with signals in the 4$^{th}$ module.
           The events in the range from 0-300~h were used for the absolute position 
           determination of the drift chambers.
         }}
 \label{elastics}
\end{figure}

In order to correct a possible misalignment of the drift chambers after 10~years 
of operation in the COSY ring
a sample of events with a single tracks was used, and the relative position of chambers 
was established as these corresponding to the minimum in the $\chi^2$ distribution of the fit
made under the assumption of a straight trajectory
of a particle through both DCs~(Fig.~\ref{detectors} upper right). The absolute
position of both DCs was established utilising the shape of the kinematical ellipse of the
elastically scattered protons~(Fig.~\ref{elastics}).
\subsection{Target}
The intersection of COSY beam and cluster target stream defines the reaction region. 
The centre of that volume is always assumed as origin 
for the reconstruction
of the trajectory of protons ejected from any point in this region. 
Therefore in order to
decrease the reconstructed momentum spread one needs to decrease the reaction region.
Such decrease
additionally lowers the 
momentum spread 
of the proton beam overlapping with the target
due to the dispersion in the target area in front of the dipole magnet.

\begin{figure}[!b]
 \begin{center}
  \epsfig{file=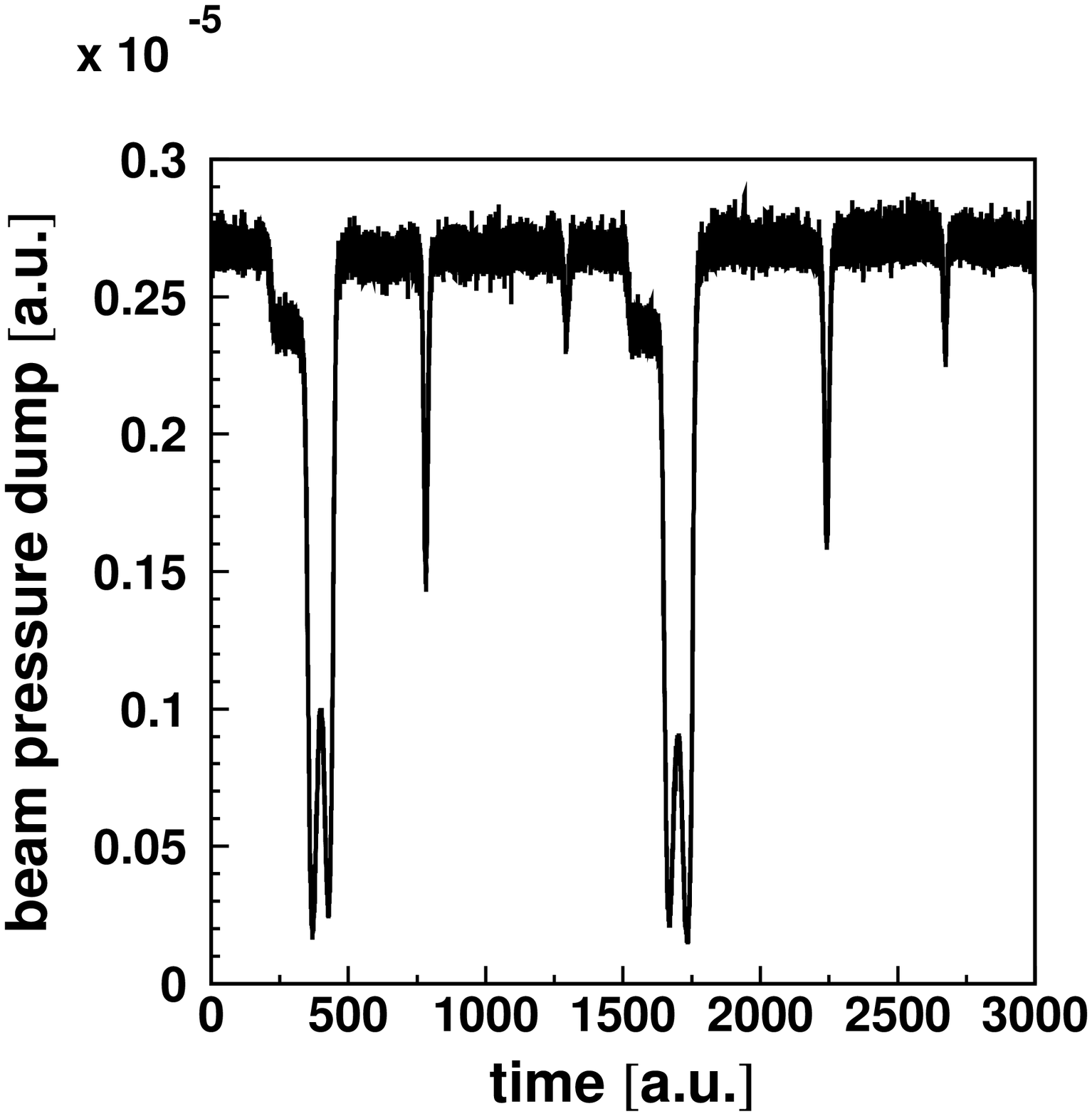,width=0.39\textwidth,angle=0}
  \epsfig{file=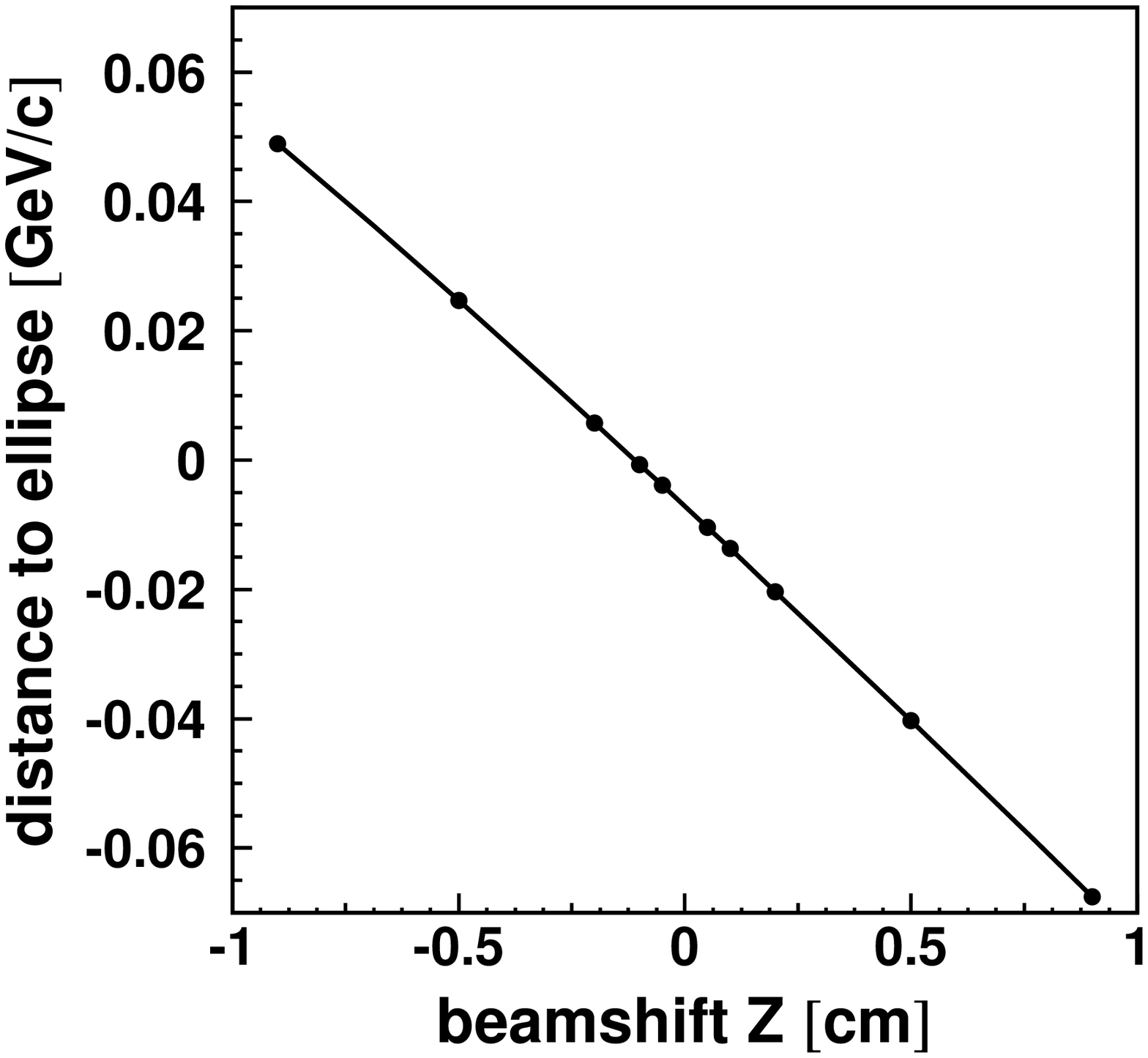,width=0.39\textwidth,angle=0}
 \end{center}
 \caption{\small{{\bf Left:}~Profile of the pressure measured during the wire device rotation.
           The observed structures depend on the position and
           size of the cluster target stream.
          {\bf Right:}~(MC)~Distance of the reconstructed kinematical ellipse of the elastically scattered protons
           to the nominal (expected) one as a function of the relative position of the target beam. 
           The observed deviation for a non-shifted target are due to the different procedures used
           during generation and reconstruction of events.
         }}
 \label{target}
\end{figure}

For the experiment described in this report
the size of the target stream was reduced by inserting an appropriately shaped aperture.
For the precize monitoring of the size of the target stream 
in the interaction region 
a diagnosis unit with wires rotating through the cluster target beam was installed, see Fig.~\ref{c11}.
When moving the wires through the cluster target beam the pressure in that region was changed and from the pressure profile the target size and position could be determined, see Fig.~\ref{target}.
As a cross-check the size of the interaction region was monitored also based on the shape of
the kinematical ellipse of the the elastically scattered protons applying the method described in article~\cite{nim}.

Several measurements with the diagnosis unit and the continuous re\-gi\-stra\-tion of 
the $pp\to pp$ reaction
allow to check the stability of the target stream as a function of time. Small fluctuations on a large time scale
were observed~(Fig.~\ref{elastics}). Such fluctuations cannot be due to
variations of the COSY beam momentum because they would require an unrealistic momentum change
ten times larger
than the typical uncertainty in the COSY settings. Any variations
of the target position in the X direction perpendicular to the proton beam would be in contradiction to
the information from the diagnosis unit.
The only reasonable solution to explain the fluctuation is a shift of the reaction
area in Z direction
(longitudinal to the proton beam) by $\approx1$~mm~(Fig.~\ref{target} right).
This variations of the position of the kinematical ellipse can be explained by fluctuations of the density inside
the target stream.

\begin{figure}[!h]
 \begin{center}
  \epsfig{file=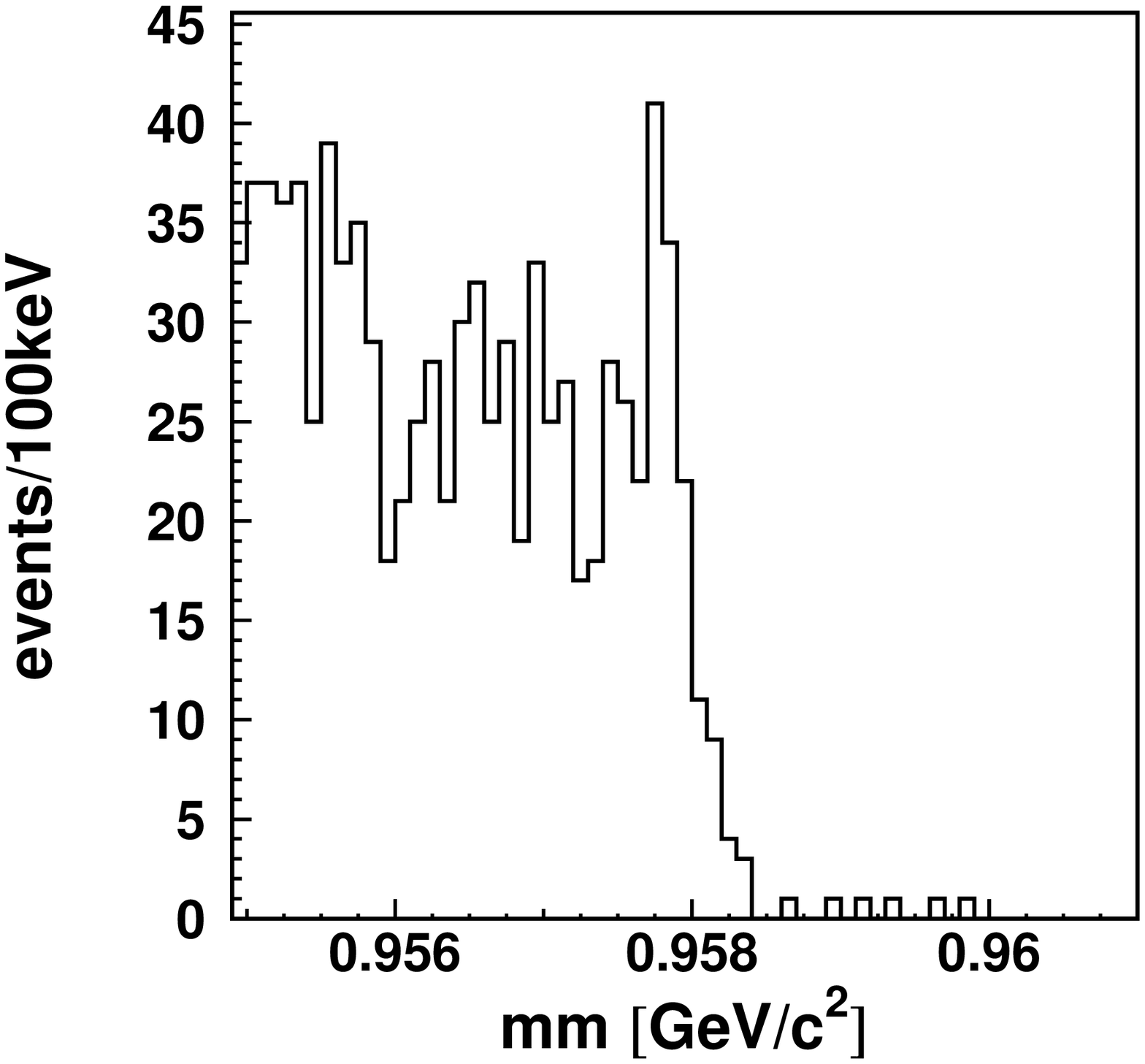,width=0.39\textwidth,angle=0}
  \epsfig{file=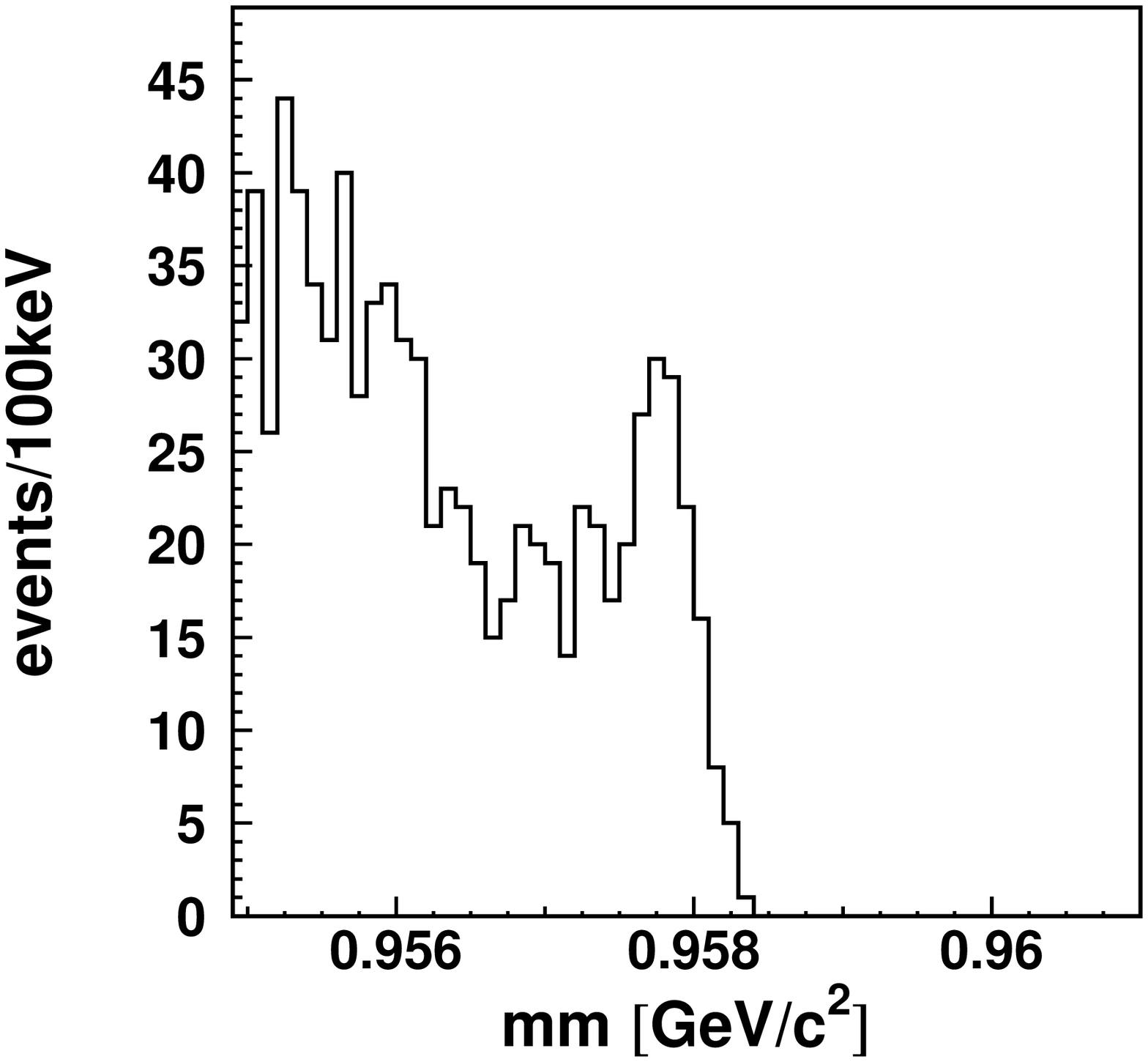,width=0.39\textwidth,angle=0}
 \end{center}
 \caption{\small{Missing mass distribution for the $pp\to ppX$ reaction for the first ({\bf left})
           and the second ({\bf right}) half of the measurement at beam momentum of 3211~MeV/c.
         }}
 \label{halfs}
\end{figure}

The observed phenomenon has a big influence on the achieved missing mass spectra as is seen
from a comparison of distributions 
obtained for first and second half of the measurement
at the beam momentum of 3211~MeV/c (see Fig.~\ref{halfs}).
%
\section{Results}
\begin{figure}[!t]
 \begin{center}
  \epsfig{file=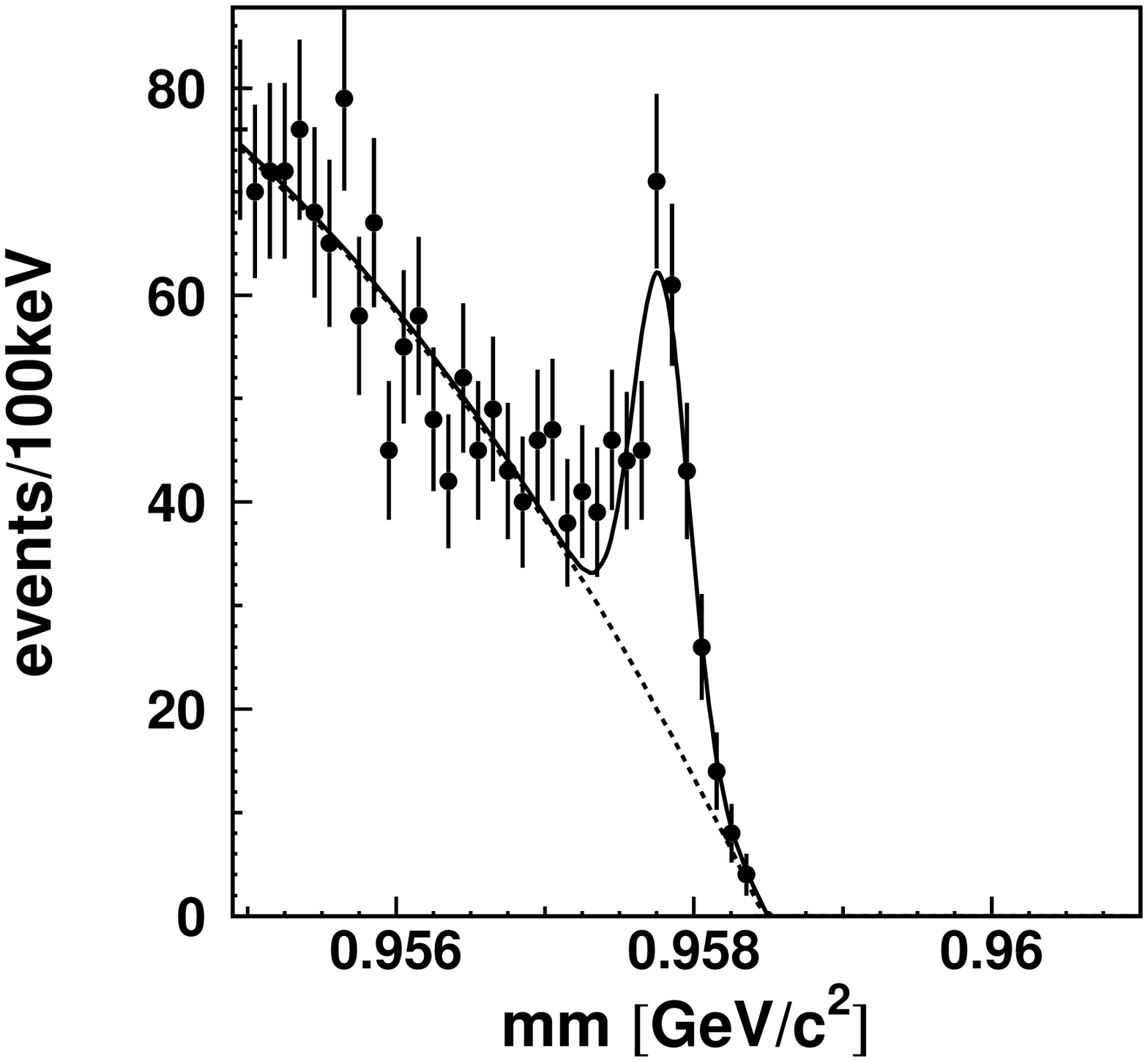,width=0.39\textwidth,angle=0}
  \epsfig{file=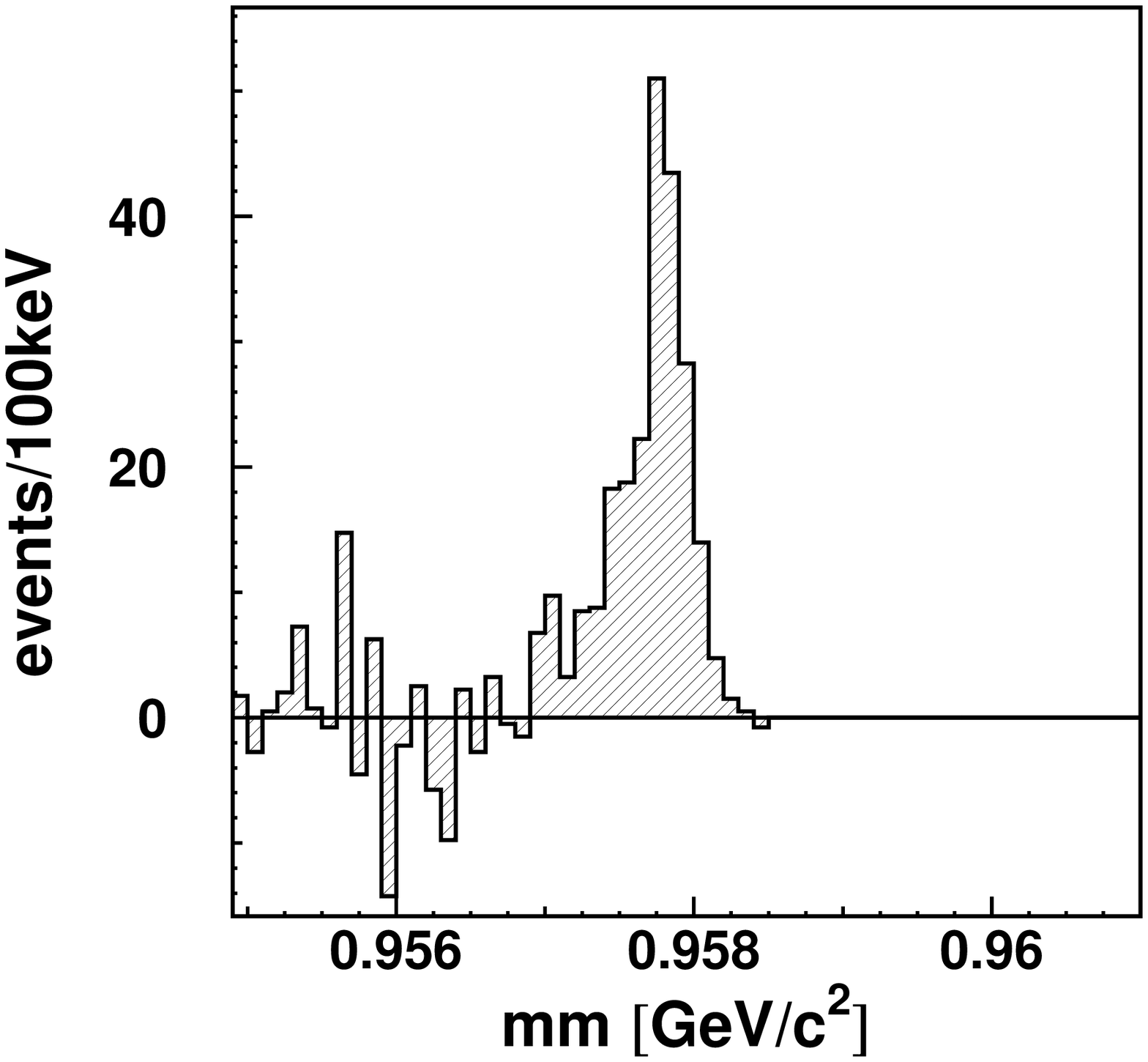,width=0.39\textwidth,angle=0}
 \end{center}
 \vspace{-5mm}
 \caption{\small{Preliminary missing mass spectra with background ({\bf left}) and
           background-corrected ({\bf right}) for 3211~MeV/c beam momentum. A 2$^{nd}$ order
           polynomial was used for the background description. For the further analysis a normalised
           background from a signal-free range in spectra from other beam momenta will be used.
           For details see~\cite{background}.
         }}
 \label{mm}
 \vspace{-7mm}
\end{figure}

As an example Fig.~\ref{mm} presents one out of five missing mass spectra obtained 
from the experiment.
The present results 
were obtained without taking into account the density variations inside the target.
%

\begin{figure}[!t]
 \begin{center}
  \epsfig{file=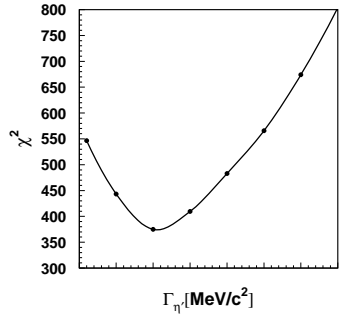,width=0.39\textwidth,angle=0}
 \end{center}
 \vspace{-5mm}
 \caption{\small{$\chi^2$ dependence on the $\Gamma_{\eta'}$ obtained from the comparison 
of the experimental and simulated missing mass 
                  spectra. For the calculation a Poisson likelihood $\chi^2$ was used 
as derived from the maximum likelihood method~\cite{hab254,hab255}
         }}
 \label{chi}
\end{figure}

For each beam momentum a set of Monte Carlo histograms 
for different assumed total width of the $\eta'$ meson
was prepared. The extraction of the value of $\Gamma_{\eta'}$ 
and of its statistical error rests on 
the simultaneous comparison 
of all five  experimental missing mass spectra 
with distributions simulated for a given value of $\Gamma_{\eta'}$.
The normalisation factors were the only free parameters in the fit.
A preliminary result of the $\chi^2$ dependence on the 
$\Gamma_{\eta'}$ is shown in  Fig.~\ref{chi}. 
The achieved statistical error in the determination of
$\Gamma_{\eta'}$ is equal to about
10~keV.
At present the absolute value of the width is under evaluation 
and it will be reported after correcting for
fluctuations of the distribution of the target density. 
%
%
\section{Acknowledgements}
The work was 
supported by the
European Community-Research Infrastructure Activity
under the FP6 program (Hadron Physics,RII3-CT-2004-506078), by
the German Research Foundation (DFG), by
the Polish Ministry of Science and Higher Education through grants
No. 3240/H03/2006/31  and 1202/DFG/2007/03,
and by the FFE grants from the Research Center J{\"u}lich.

\end{document}